\documentclass[aps]{revtex4}
\usepackage{epsfig}
\usepackage{graphicx}
\usepackage{amsmath,amssymb,color}
\usepackage[english]{babel}

\parskip=\medskipamount

\newcommand{\mtrx}[1]{{\mathbf{#1}}}

\newcommand{\lp}{\left(}
\newcommand{\rp}{\right)}

\newcommand{\eq}[1]{(\ref{#1})}
\newcommand{\fig}[1]{Fig.\ref{#1}}

\newcommand{\be}{\begin{equation}}
\newcommand{\ee}{\end{equation}}

\begin{document}

\title{Towards a robust algorithm to determine topological domains from colocalization data}

\author{
  Alexander P. Moscalets$^{1,2,3}$,
  Leonid I. Nazarov$^{4}$,
  and
  Mikhail V. Tamm$^{1,4}$
}

\affiliation{$^1${Department of Applied Mathematics, National Research University Higher School of Economics, 101000, Moscow, Russia}\\
$^2$ {A.N. Nesmeyanov Institute of Organoelement Compounds of the Russian Academy of Sciences, 119991, Moscow, Russia}\\
$^3$ {Chemistry Department, Moscow State University, 119991, Moscow, Russia}\\
$^4$ {Physics Department, Moscow State University, 119991, Moscow, Russia}}


\begin{abstract}

One of the most important tasks in understanding  the complex spatial organization of the genome consists in extracting information about this spatial organization, the function and structure of chromatin topological domains from existing experimental data, in particular, from genome colocalization (Hi-C) matrices. Here we present an algorithm allowing to reveal the underlying hierarchical domain structure of a polymer conformation from analyzing the modularity of colocalization matrices. We also test this algorithm on several model polymer structures: equilibrium globules, random fractal globules and regular fractal (Peano) conformations. We define what we call a spectrum of cluster borders, and show that these spectra behave strikingly differently for equilibrium and fractal conformations, allowing us to suggest an additional criterion to identify fractal polymer conformations. 

\end{abstract}
\maketitle

\section{Introduction}

The question of how an astonishingly long DNA chain consisting of $10^{9}$ basepairs or even more folds into a compact state within a small volume of cell nucleus is still intriguing and not completely resolved. Indeed, not only is the packing compact which would have been easily achievable if DNA was in a so-called equilibrium globule state \cite{deGennes, GrKhokh, RubColby}, but it is also capable to function biologically in a meaningful way. This biological function enforces many very specific and clearly non-equilibrium features including the existence of distinct chromosome territories and topological domains (TADs) within single chromosome, easy unentanglement of chromosomes and chromosome parts (needed in preparation to mitosis, and during transcription), and ability of different parts of the genome to find each other in space strikingly fast in e.g. so-called promoter-enhancer interactions. The concrete mechanism which stabilizes these features is not yet completely understood. The main candidates for this stabilization is the so-called model of non-equilibrium fractal globule \cite{GNS, grosbconj, rosaeveraers, Lieberman, MirnyChrom, grosbreview, GrosbSoft, RosaEveraersnew, NazarovSoft, TammPRL, BunKardar}, as well as various models accentuating the formation of saturating bonds between the fragments of chromatin \cite{loop1, loop2, loop3, loop4, loop5, pnas, LongDiff1, LongDiff2}.

On the experimental side, the high-resolution data concerning the spatial organization of the genome is mostly due to the development of the genome-wide chromosome conformation capture (so called Hi-C) method \cite{dekker, Lieberman, naumova} which allows to obtain the {\it colocalization} matrices of the genome packing, containing information about which particular genome fragments are {\it closed to each other} in space. It is from the statistical analysis of these matrices that the authors of  \cite{Lieberman, MirnyChrom} deduced that the fractal globule state is a suitable candidate to describe chromosome packing. Originally, the Hi-C matrices represent the colocalization data averaged over many cells, however, recently a significant progress was reported in obtaining single-cell Hi-C maps \cite{nagano2013single}.

Note therefore, that the data available include not the information about full spatial organization of the genome, but just on the parts of the genome which are spatially close to each other. It is, therefore, of great interest to develop methods extracting as much information as possible from such a dataset. In particular, it is a challenging question whether one can recover the information about the exact structure of TADs from colocalization data. It seems in principle reasonable to believe that such extraction is possible: indeed, parts of a chromosome belonging to a single TAD are spatially compact and should therefore more often find themselves in a close proximity with each other.

In this paper we suggest an algorithm, based on the methods of complex network theory \cite{PhysRevE.70.056131, PhysRevE.84.066122, doi:10.1142/S0218127412501714, 1367-2630-10-5-053039} which allows to reveal a hierarchical TAD structure of a polymer conformation if it does exist, and check the applicability of this algorithm on several model polymer conformations. In what follows we discuss this algorithm for a particular case of a single-conformation colocalization matrix whose elements are `0's and `1's depending on whether the two corresponding monomers are spatially adjacent or not. The generalization for a more experimentally typical case of Hi-C maps {\it averaged over} many conformations is absolutely straightforward. Note, however, that the significance of resulting community structure is not ensured: indeed, as we argue elsewhere \cite{NazarovSoft} the experimental Hi-C maps are in fact averaged over many substantially different folding conformations, and one expects therefore the community structure of the average to be significantly less rich than the community structure of the individual conformations. We hope, however, that ({\it i}) further advances in the single-cell Hi-C mapping techniques will allow in-depth analysis of the community structure of individual genome conformations, and ({\it ii}) that comparison of the community structures corresponding to individual and averaged Hi-C maps may shed light on which characteristics of genome folding are conserved from realization to realization and which ones are variable.

\section{Model description}

It is feasible to construct a mapping between any given configuration of a polymer chain of $N$ monomers and a colocalization matrix $\mtrx{W}$ with matrix elements $w_{ij}$ equal to 1 if $i$-th and $j$-th monomers are close to each other in space (i.e., if the spatial distance $r_{ij}$ is less than some cut-off value $a$ which here and below is presumed to be of order of the monomer-monomer bond length) and 0 otherwise. Our aim is to extract the information about topological domain structure of the chain conformation from this matrix $\mtrx{W}$, and to estimate if this domain structure is stable (in some sense to be determined later), and if it possesses ultrametric properties (i.e., consists of sequentially nested domains of smaller and smaller size) expected from the structure of fractal globule (see, e.g., \cite{GNS, MirnyChrom, NazarovSoft, BunKardar} for the discussion of ultrametricity in the context of fractal globules). In what follows we develop a method allowing us to do that and show the results obtained by this method on several test polymer configurations. Application of this method to real experimental data goes beyond the scope of this paper and will be provided elsewhere.
 
The analysis of matrix  $\mtrx{W}$ starts from the notion that it could be reinterpreted as an adjacency matrix of some complex graph (network), which allows us to use the community detection methods developed in the complex network theory \cite{FortunatoPhysicsReports}. Indeed, it seems natural to assume that monomers belonging to the same TAD will more often be in spatial proximity to each other in space than monomers belonging to different TADs, and therefore a community of comparatively well-connected nodes in the network theory sense can be a good proxy to a real topological domain. It is known, however, that the problem of optimal division of a network into a set of communities (comparatively well connected clusters) is ill-posed and does not have a single definite answer; there exist numerous techniques based on spectral properties of adjacency matrix \cite{Newman06062006}, non-backtracking matrix \cite{Krzakala24122013}, synchronization in networks \cite{ArenasPhysicsReports}. In what follows we employ the modified modularity-optimization method as developed in  \cite{PhysRevE.70.056131, PhysRevE.84.066122, doi:10.1142/S0218127412501714, 1367-2630-10-5-053039} which we find most appropriate to our needs as it very naturally allows to look into the community structure of networks on different scales.

\begin{figure}
\epsfig{file=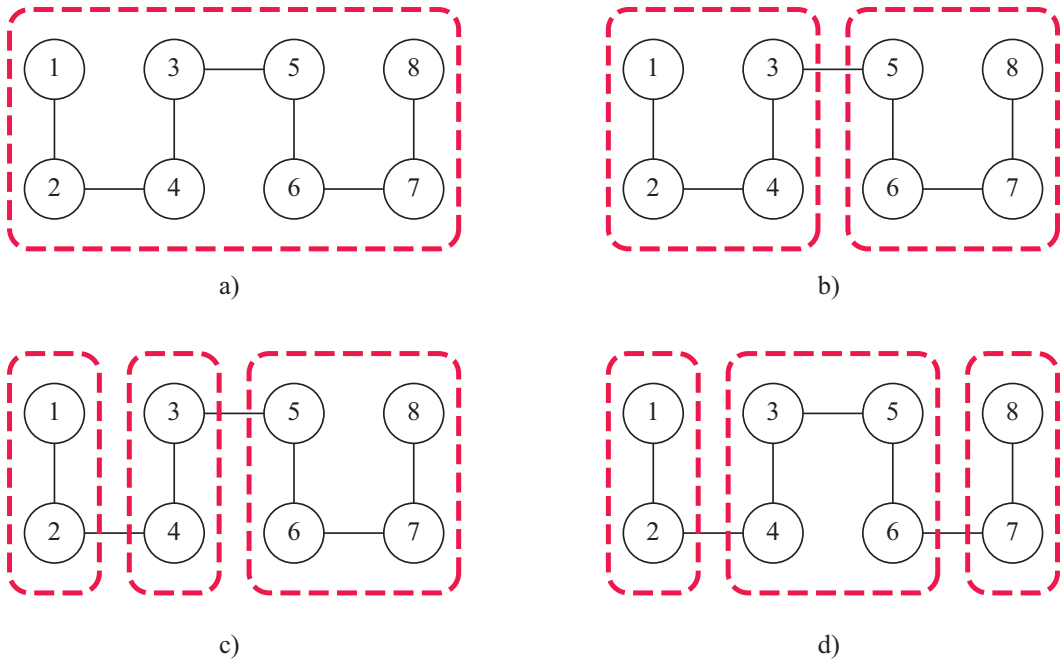, width=14 cm} \caption{ A toy model of a network consisting of 8 nodes and its possible separations into clusters. See explanation in the main text for details. }
\label{FigEx1}
\end{figure}

Consider a partitioning of an $N$-node network into a set of $k$ clusters (see \fig{FigEx1} below for a toy example of such partitioning). Any such partitioning can be described by a matrix $\mtrx{C}$ of size $N \times k$ with $c_{i\alpha}=1$ if $i$-th element of the network belongs to a $\alpha$-th cluster and 0 otherwise. Naturally, one assumes $\sum_{\alpha} c_{i \alpha} = 1$ for any $i$ (i.e., each node belongs to one and only one cluster). Then, according to \cite{PhysRevE.70.056131} the modularity $Q[\mtrx{W},\mtrx{C}]$ of such a partition is defined as
\begin{equation}
Q[\mtrx{W},\mtrx{C}]=\frac{1}{2w} \sum_{i,j}\lp w_{ij}-\frac{w_i w_j}{2w}\rp \sum_{\alpha}{c_{i\alpha}c_{j\alpha}},
\label{e1}
\end{equation}
where $w_{ij}$ are the elements of $\mtrx{W}$, $w_i=\sum_j w_{ij}$ is the total number of neighbors of $i$-th node (also called `strength' of the node), $w=\sum_{ij} w_{ij}/2$ is the total number of links (strength) of the network; the sum over $\alpha$ in \eq{e1} equals 1 if $i$-th and $j$-th monomers belong to the same cluster, and 0 otherwise. The original modularity-based community detection algorithm demands to maximize the functional $Q$ with respect to $k$ and $\mtrx{C}$, the corresponding maximizing separation is then considered to be optimal. Thus defined method	 of modularity optimization is known to have two important drawbacks. First, it is known to have a so-called resolution limit, so that it is impossible to find any clusters of size less than $\sim \sqrt{2w}$ correctly \cite{PhysRevE.84.066122}. Second, the total number of possible partitions of a network into clusters is exponentially large in $N$ making the search of an optimal partition an NP-complete problem (i.e., a problem whose shortest possible solution time grows exponentially with $N$, see \cite{sethna, mez_mont} for the introduction to this concept).

There exists, however, a possibility to circumvent the first problem. Following \cite{doi:10.1142/S0218127412501714, 1367-2630-10-5-053039} introduce the so-called resistance parameter $r$, i.e., introduce a modified adjacency matrix  $\bar{\mtrx{W}}(r)$ whose matrix elements equal to $\bar{w}_{ij}=w_{ij}$ for $i \neq j$, and $\bar{w}_{ii}=r$. In terms of the underlying network this corresponds to adding self-loops with weight (strength) $r$, which is, generally speaking, non-integer and can be even negative. Now, proceed with the optimization of the modularity functional \eqref{e1} for this new matrix $\bar{\mtrx{W}}(r)$ (note that the definition of modularity never relies on the elements of adjacency matrix being Boolean variables, indeed, it was originally introduced for weighted networks). The larger the resistance parameter $r,$ the more are nodes coupled to themselves as compared to other surrounding nodes. As a result of that, as shown in\cite{1367-2630-10-5-053039} smaller and smaller clusters get determined. Indeed, if $r$ is larger than some network-dependent critical value $r_{\max}$ the optimal partition is one separating the network into $N$ clusters consisting of 1 monomer each. On the other hand side, if $r$ is less than some (once again, network dependent) $r_{\min}$ which is usually negative, the optimal partition consist of a single cluster covering the whole network.

As for the second drawback of usual modularity algorithm, i.e. the exponential increase of the possible number of separations with the size of the system, it also can be circumvented in this particular case. Indeed, remind that we originally defined the adjacency matrix $\mtrx{W}$ as a colocalization matrix of some {\it polymer} configuration. That is to say, the monomers of the network are naturally numbered along the chain, and monomers close along the chain are automatically close in space due to the connectivity of a polymer. This allows us to postulate by definition that we only consider clustering partitions which separate a chain in fragments which are adjacent along the chain, i.e. if $i$-th and $j$-th monomer belong to the same cluster, than any $k$-th monomer with $i<k<j$ belong to the same cluster as well. Note, that such definition of clusters is in accordance with how topological domains are usually understood in polymer and biophysical literature: parts of the chain that are close  \emph{both} along the chain and in real space. Simultaneously, it is easy to see that such a restriction on possible partitioning reduces their overall number from exponential in $N$ to quadratic in $N$, allowing us to produce a rather fast deterministic partitioning algorithm, which we realized in {\sc Fortran 95}.

For the purposes of our work, it is instructive to consider how the resulting partitioning (community structure) {\it evolves} with the change of $r$. Indeed, if one considers a completely random Erd{\H{o}}s--Renyi network, one would expect that at $r_{\min}$  the network is separated into two clusters of roughly same size, then as the value of resistance reaches some $r_2>r_{\min}$ it separates into three clusters of, once again, roughly same size, than at $r_3>r_2$  into four clusters, etc. The important thing is that in the absence of any underlining structure of the network one expects the cluster boundaries to be essentially uncorrelated: every time the number of clusters increases by one and \emph{all} cluster boundaries rearrange. On the other hand, if the networks has an underlying structure of hierarchically organized TADs, one expects the increase of $r$ to cause not a complete rearrangement of cluster structure, but a decomposition of already existing clusters into smaller parts with boundaries of larger clusters remaining essentially stable. 
In order to quantify this qualitative notion we introduce here the concept of a \emph{spectrum of borders}.

Let $H_i(r)$ ($i=1,\ldots ,N-1$) be a Heaviside step function indicating whether the bond between $i$-th and $(i+1)$-th monomer is a border of a cluster at given $r$ (that is to say $H_i(r)=1$ if the said bond is a boundary, and $H_i(r)=0$ otherwise). Then the total fraction of the range $(r_{\min},r_{\max})$ when a given bond is a border of a cluster is given by
\begin{equation}
f_i=(r_{\max}-r_{\min})^{-1}\int_{r_{\min}}^{r_{\max}} H_i (r)\,dr.
\label{e3}
\end{equation}
The higher $f_i$ is, the more stable is the cluster border at $(i, i+1)$ bond. The whole set of $f_i$s  for all $i$ is what we call a spectrum of borders (SoB). Inspecting such a spectrum one should be able to estimate how stable the cluster structure of the network is, and what exactly is the natural hierarchical domain structure of a network, if any. If there is no well-defined clusters in a network, the SoB is a more or less uniform distribution of lines, while if there is a small number of $f_i$ significantly exceeding the rest, it signifies a network with stable and well-defined cluster structure.

\begin{figure}
\epsfig{file=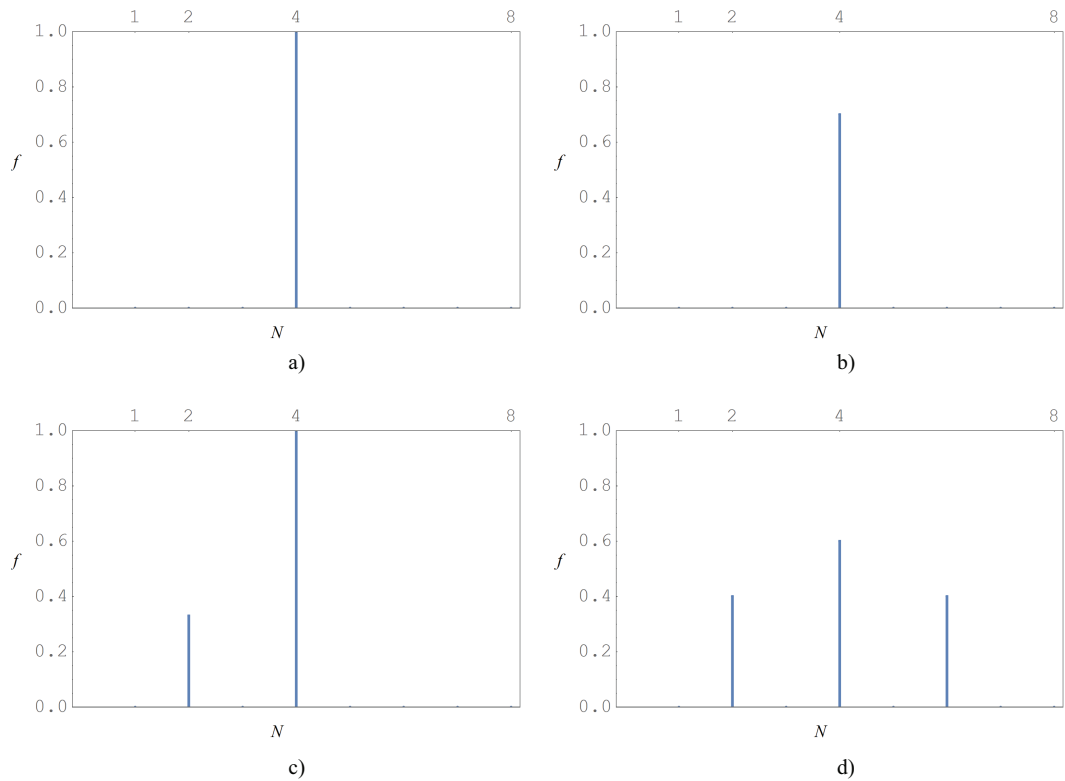, width=14 cm} \caption{Spectra of borders for a toy eight-nodes network shown in \fig{FigEx1}. For the explanation of differences between the boxes see the main text. }
\label{FigEx2}
\end{figure}

To clarify this definition, let us start with a toy example. Consider a model network with 8 consequentially numbered nodes (see \fig{FigEx1}a) which we study in the interval between some fixed $r_{\min}$ and $r_{\max}$. Assume first that along the whole  interval it is partitioned into exactly two clusters with the border at node 4 (\fig{FigEx1}b). Then
\begin{equation}
f_i=\begin{cases}
1, & \text{for $i=4;$} \\
0, & \text{otherwise.}
\end{cases}
\end{equation} 
This means that in the whole range of $r$ there is only one border between two clusters and this border does not shift to the left or right, thus the clusters always contain exactly 4 nodes each (i.e., if partitioning shown in \fig{FigEx1}b holds for any $r$, than the SoB is the one shown in \fig{FigEx2}a). 
If there appears a region within which these clusters vanish and optimal partitioning consists of just one cluster (i.e., \fig{FigEx1}b for some $r$s and \fig{FigEx1}a for some others), than the amplitude $f_{4}$ becomes less than 1 (\fig{FigEx2}b).

Suppose now, that above certain value $r_2\in (r_{\min},r_{\max})$ the first of the two clusters splits into 2 smaller ones of size 2 each (see \fig{FigEx1}c). This will result in the emergence of new line of smaller amplitude in the SoB at node 2 (\fig{FigEx2}c):
\begin{equation}
f_i=\begin{cases}
1, & \text{for $i=4;$} \\
f,& \text{for $i=2;$}\\
0, & \text{otherwise,}
\end{cases}
\end{equation} 
where
\begin{equation}
f = \frac{r_{\max}-r_2}{r_{\max} - r_{\min}}, \quad 0<f<1.
\end{equation}
A different possibility is that at $r_2\in (r_{\min},r_{\max})$ the border at node 4 disappears and new set of three clusters consisting of, say, nodes $(1,2)$, $(3,4,5,6)$, and $(7,8)$ arise (see \fig{FigEx1}d). Then the set of amplitudes in the SoB will read
\begin{equation}
f_i=\begin{cases}
1-f, & \text{for $i=4;$} \\
f,& \text{for $i=2,6;$}\\
0, & \text{otherwise,}
\end{cases}
\end{equation} 
The corresponding spectra of borders are shown in  \fig{FigEx2}c and  \fig{FigEx2}d, respectively.

Thus, in accordance with what have been said above, if the borders of clusters  dangle more or less randomly along the chain, the resulting SoB consists of a number of low peaks of roughly equal height, and such a behavior is characteristic for networks with labile, fuzzy domain structure. On the contrary, the changes when the existing large cluster splits into smaller ones without changing its outer borders give rise to a series of peaks which significantly differ in height. Such behavior seems to be characteristic of networks with well-defined hierarchical domain structure.

\section{Results and discussions}

To check the aptitude of the described approach we applied it to series of adjacency matrices of several model polymer conformations. In particular, we studied, (i) a completely deterministic Peano curve which is the simplest possible proxy of a fractal globule conformation with most prominent hierarchical self-similar cluster organization possible, (ii) an equilibrium conformation of a Gaussian polymer globule, (iii) a random fractal globule conformation obtained by the conformation-dependent polymerization. 

The first two conformations are more or less standard, while the third, suggested originally in \cite{TammPRL} (the algorithm is briefly outlined below, and we address the reader to the supplementary materials of that paper for full details about this algorithm) is, in our opinion, one of the best existing candidates to represent the generic metastable fractal globule state. It shows significant stability in dynamic computer simulations and its statistical characteristics are very similar to what is obtained in other fractal-globule generating algorithms. Therefore, the main aim of our test is to show that random fractal globule conformations could be robustly and reproducibly distinguished from equilibrium Gaussian ones based on their SoBs.

The length of polymer under consideration was $N=4096$ for Peano curve and $N=10000$ for random fractal and equilibrium conformations. All chain configurations were generated on cubic lattice. For generation of the fractal globule the following rules were used. The monomers are added sequentially to the chain, with a new monomer added to one of the 6  nodes of a lattice adjacent to the endchain monomer with probabilities $p_i, \, i=1,\ldots ,6$ to go in each direction equal to: 
\begin{equation}
p_i = \mathcal{N}^{-1}\begin{cases}
10^{-6},& \text{if $i$-th site is visited},\\
1+10^4 n_i,& \text{if $i$-th site is not visited},
\end{cases}
\end{equation}
where $\mathcal{N}=\sum_i p_i$ is a normalizing constant, and $n_i$ is the number of occupied sites within a unit sphere centered at site $i$. The significant difference with \cite{TammPRL} is that the fractal globules were obtained in a free volume without periodic boundary conditions, which allowed to obtain conformations with very developed surface.
Equilibrium globule was obtained as random walk within a sphere of radius $R=1.6\lp \frac{3N}{4\pi}\rp^{1/3}$ and reflecting surface. In case of trapped configurations the chain end can go back through randomly chosen, among already visited, site. 

The elements of adjacency matrix $w_{ij}$ equal to 1 (monomers are considered to be neighbors) if they are separated by distance $r_{ij}\leq a =\sqrt{3}$ lattice units. In our analysis we somewhat arbitrarily have chosen $r_{\min} = -20$ and $r_{\max} = 80$. While $r_{\min}=-20$ roughly corresponds to the point where the first partitioning of a network into two clusters takes place for the networks under study, the chosen value of  $r_{\max} = 80$ is significantly smaller than the natural limit defined above. That is to say, even for largest $r$ under consideration, our networks are quite far from being separated into $N$ single node clusters.  This choice was dictated mostly by the saving of CPU modeling time. The calculations presented below thus grasp the essential behavior of the SoB for large clusters and the most stable domain walls, which are, in our opinion, of the most interest. Indeed, small-scale clustering structure of our test configurations can be significantly plagued by the underlying discrete lattice. 
However, we perform a check of the robustness of suggested algorithm with respect to the choice of $r_{\max}$ (see below).

The application of the proposed algorithm to Peano curve reveals hierarchically nested clustering, as expected. At values $r<-20.7$ there is the only community with 4096 nodes. At $r\approx -20.7$ two clusters, consisting of 2048 nodes each, appear. With further increase of $r$ each of these clusters splits exactly into 2 half-size clusters, but the border at node $i=2048$ remains at place, then each of the four clusters splits into halves, etc, all the way until $r=r_{\max}.$ Therefore, $f_{2048}$ is the highest peak in SoB for Peano curve, followed by $f_{1024}$ and $f_{3072}$ , etc. We get finally a SoB consisting of an hierarchical set of equidistant peaks as shown in Figure 2. Note that smaller clusters positioned deep inside the globule and on its outer surface behave somewhat differently, which explains why starting from the third generation the values of peaks of the same generation are not exactly equal.  

\begin{figure}
\epsfig{file=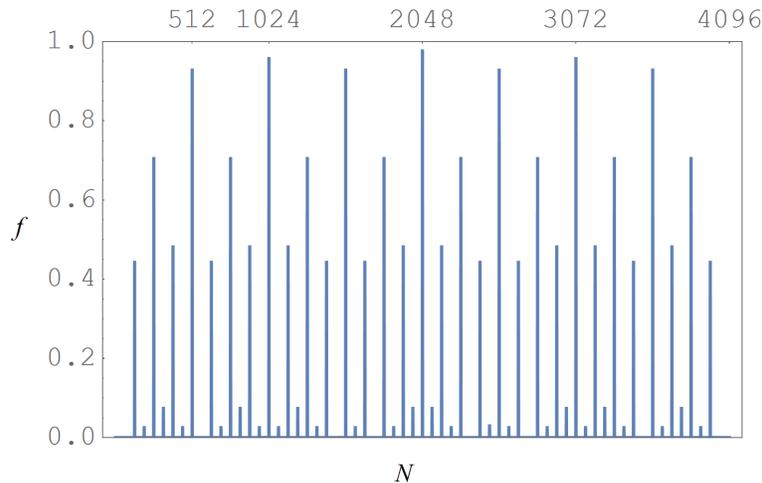, width=10 cm} \caption{Spectrum of borders (magnitude $f$ vs node number $N$) obtained for complex network corresponding to Peano curve. }
\label{Fig2}
\end{figure}

The \fig{Fig2} shows six levels of hierarchical organization of Peano curve,  each larger cluster (consisting of the nodes between two largest peaks) having a fully deterministic internal structure. For example, the second level cluster, consisting of the nodes 2048 to 3072, has two domains (divided by line at node 2560) of  512 nodes each. The fourth level of hierarchy is organized by eight peaks on nodes $256+512n$ ($n=0,\ldots ,7$) with those of higher level, which make it a total of 16 clusters. Such a picture of discrete lines with gradually lowering heights at each level is a qualitative representation of regularly fractal, self-similar polymer globule.

\begin{figure}
\epsfig{file=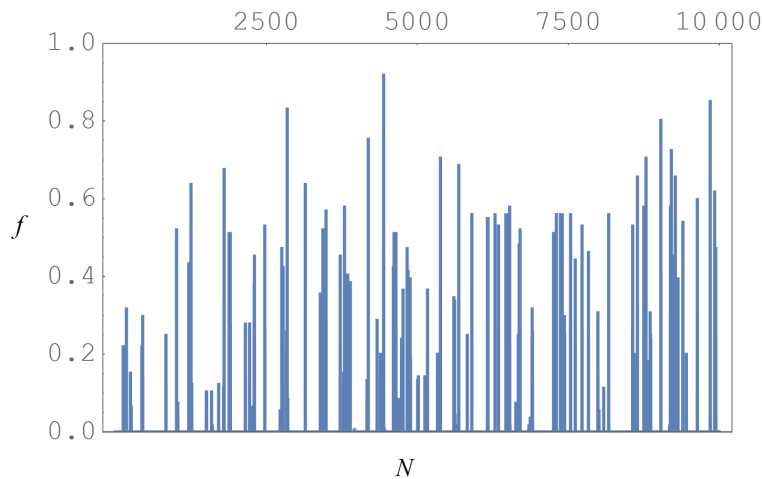, width=10 cm} \caption{ Spectrum of borders (magnitude $f$ vs node number $N$) obtained for complex network corresponding to fractal (sticky) globule.}
\label{Fig3}
\end{figure}

Consider now a random fractal globule. Clearly, it does not have such a regular fully symmetric domain structure (see \fig{Fig3}), whatever one may notice that the peaks in the structure are once again very widely distributed: there are some relatively very high peaks (e.g., ones at nodes 4441, 9849, 2844 reach well above 0.8) and the general structure of the SoB is very widely disperse.

\begin{figure}
\epsfig{file=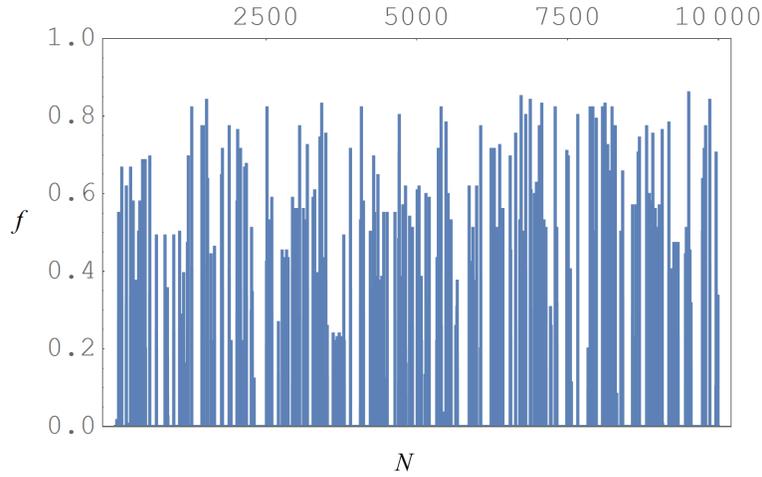, width=10 cm} \caption{ Spectrum of borders (magnitude $f$ vs node number $N$) obtained for complex network corresponding to equilibrium globule.}
\label{Fig4}
\end{figure}

Contrary to that,  the SoB for an equilibrium globule (\fig{Fig4}) is essentially a dense forest of peaks, peaks of any height seem to appear with equal probability without any gaps between higher and lower peaks. Such a picture tells us that almost each of the node has been the border of any cluster, or in other words, as it was described above, on magnifying the network the cluster structure rearranges completely with changing $r$, confirming thus that an equilibrium globule does not have a well-defined cluster structure.

\begin{figure}
\epsfig{file=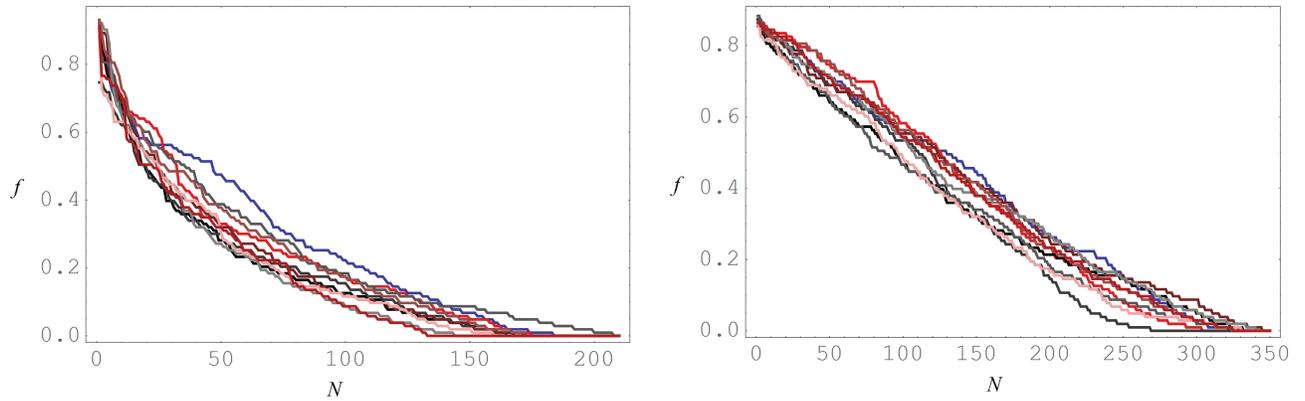, width=17 cm} \caption{Ordered amplitudes $f$ from spectrum of borders for 10 different realizations of fractal (left) and equilibrium (right) globules.}
\label{Fig5}
\end{figure}

In order to make the difference between figures \fig{Fig3} and \fig{Fig4} more clear, it is instructive to reorder peaks of the SoB in descending order, constructing thus an {\it ordered SoB}, where the highest peak is renamed $f_1$, the second highest $f_2$, and so on. \fig{Fig5} shows the envelope lines of ordered SoBs for 10 different random fractal realizations (all of them cluster in the right bottom of the picture) and 10 different equilibrium globule conformations (which cluster in the top left). One sees immediately that spectra of fractal and equilibrium globules are clearly distinguishable: spectra of equilibrium conformations show approximately linear descent, confirming that in such conformation borders of all intensities occur with roughly the same probability, while fractal globule has a clear cluster structure with a small number of very strong borders dominating over the rest.

\begin{figure}
\epsfig{file=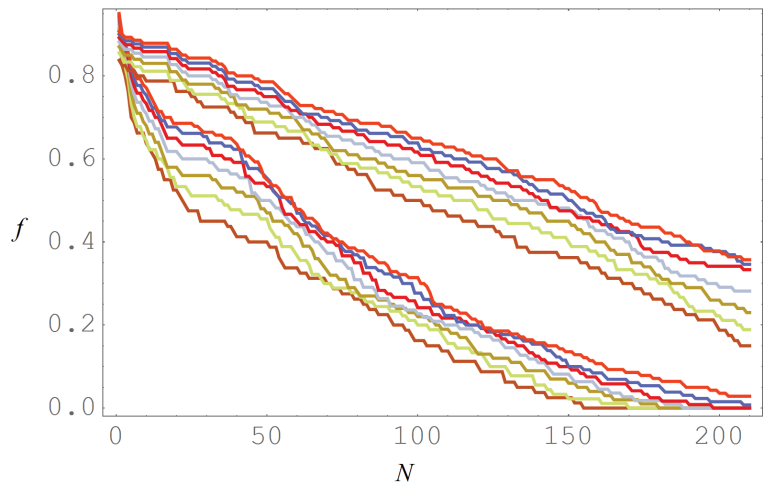, width=10 cm} \caption{Comparison of averaged ordered amplitudes $f$ of fractal (bottom) and equilibrium (top) globules at different values of parameter $r_{\max}$ chosen from interval 80 to 140. Curves of the same color are obtained at the same values of $r_{\max}$.}
\label{Fig1}
\end{figure}

Moreover, in order to check if this distinction between two classes of conformations is robust with respect to change in $r_{\max}$, we have plotted a series of ordered SoBs for a single random fractal and a single equilibrium conformation but different values of $r_{\max}$ varying from 80 to 140, see \fig{Fig1}. It is clear from the Figure that the changes of the curves, although clearly visible, do not change the general result: once again the curves for fractal and equilibrium conformations are clearly distinguishable.

\begin{figure}
\epsfig{file=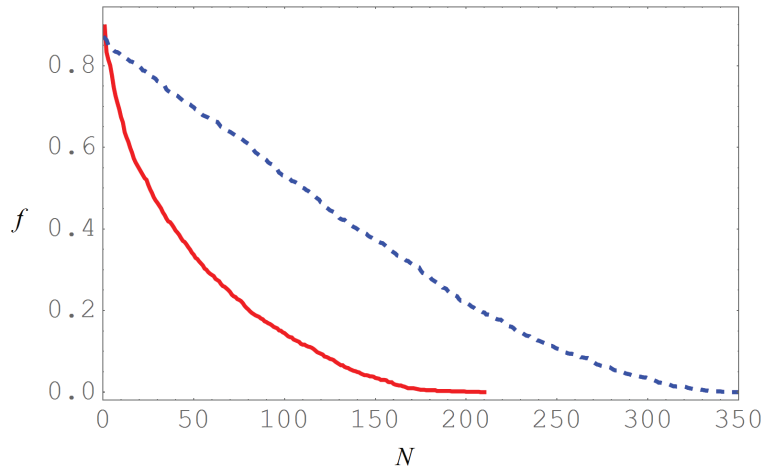, width=10 cm} \caption{Comparison of averaged ordered amplitudes $f$ of fractal (red solid) and equilibrium (blue dashed) globules.}
\label{Fig6}
\end{figure}

\fig{Fig6} shows the average shape of ordered SoBs, obtained from averaging the curves shown in \fig{Fig5} over 10 different conformations of the same class. Clearly, the spectra have very different shape, the curve corresponding to the equilibrium globule can be fitted as a straight line, whereas the case of fractal globules highly non-linear with a steep descent at the beginning. We are now working on a possibility to construct theoretical explanations for the shapes of the obtained curves, corresponding results will be presented elsewhere.

We expect  that the concept of the ordered spectrum of borders introduced above will provide an additional tool to differentiate between different possible conformation types of the polymer molecules, while the exact positions of the most strong borders will provide information about the position of TAD borders in real chromosome conformations.

In conclusion, let us discuss once again the possible applicability of the algorithm suggested in this paper to the experimental  Hi-C maps, which are usually averaged over millions of cell with, generally speaking, different folding structures. Such maps are in fact symmetric matrices with non-negative non-integer elements, and they can be naturally interpreted as adjacency matrices of \emph{weighted} complex networks. The algorithm we suggest does not at any point rely on the fact that the network under consideration is unweighted, and thus can be applied without change to this averaged Hi-C maps. There is, however, a more subtle question. It is not completely clear how much of the original hierarchy of the chain packing is conserved from realization to realization. We expect, and plan to check elsewhere, that those parts of the TAD structure which are repeted in all (or most) cells, should be accessible from analysing community structure of averaged Hi-C maps. On the contrary, those parts which are different from cell to cell should be observable only from the single-cell Hi-C maps (provided the corresponding experimental techniques will progress) but we expect them to be smeared over in the averaged maps.  We expect that the comarison of the community structures of the averaged and single-cell Hi-C maps can elucidate (especially coupled with the progress of the single-cell Hi-C mapping techniques) the question of how stable individual TADs are from cell to cell (see, e.g. 
\cite{NazarovSoft} for further discussion of this subject).
  
\section*{Acknowledgments}
Authors are grateful to S.~K.~Nechaev, L.~Mirny, A.~V.~Chertovich, and P.~Kos for fruitful discussions and to A.~Cherstvy for his useful comments on the text of the manuscript. The work is partially supported by the RFBR grant 14-03-00825 and by the Higher School of Economy program for basic research.

\section*{Conflict of interest}
All authors declare no conflicts of interest in this paper.

\end{document}